\UseRawInputEncoding
\documentclass[twocolumn,amsmath,amssymb,aps,pra,superscriptaddress]{revtex4-2}

\usepackage{mypackages}
\usepackage{tikz}
\usetikzlibrary{shapes.geometric, arrows, positioning}

\tikzstyle{process} = [rectangle, rounded corners, minimum width=3.5cm, minimum height=1.2cm, text centered, draw=black, align=center,font=\Huge]
\tikzstyle{arrow} = [thick,->,>=stealth, line width=0.8mm, scale=3]

\usepackage{amscd}
\usepackage{textcomp}
\usepackage[thinlines]{easybmat}
\usepackage{multirow,bigdelim}
\usepackage{adjustbox}

\usepackage{booktabs}
\usepackage{layouts}

\usepackage{pgf}
\usepackage{import}
\usepackage{xcolor}

\usepackage[
    font={small}, 
    labelfont=bf,
    justification=justified,
    singlelinecheck=false
]{caption}

\usepackage{ragged2e}

\begin{document}

\title{Colour Codes Reach Surface Code Performance using Vibe Decoding}%

\author{Stergios Koutsioumpas}
\email{skoutsio@ed.ac.uk}
\affiliation{School of Informatics, The University of Edinburgh, United Kingdom}
\affiliation{Department of Physics \& Astronomy, University College London, United Kingdom}
\author{Tamas Noszko}
\affiliation{School of Informatics, The University of Edinburgh, United Kingdom}
\author{Hasan Sayginel}
\affiliation{Department of Physics \& Astronomy, University College London, United Kingdom}
\author{Mark Webster}
\affiliation{Department of Physics \& Astronomy, University College London, United Kingdom}
\author{Joschka Roffe}
\email{joschka@roffe.eu}
\affiliation{School of Informatics, The University of Edinburgh, United Kingdom}

\begin{abstract}
Two-dimensional quantum colour codes hold significant promise for quantum error correction, offering advantages such as planar connectivity and low overhead logical gates. Despite their theoretical appeal, the practical deployment of these codes faces challenges due to complex decoding requirements compared to surface codes. This paper introduces vibe decoding which, for the first time, brings colour code performance on par with the surface code under practical decoding. Our approach leverages an ensemble of belief propagation decoders---each executing a distinct serial message passing schedule---combined with localised statistics post-processing. We refer to this combined protocol as VibeLSD.
The VibeLSD decoder is highly versatile: our numerical results show it outperforms all practical existing colour code decoders across various syndrome extraction schemes, noise models, and error rates.
By estimating qubit footprints through quantum memory simulations, we show that colour codes can operate with overhead that is comparable to, and in some cases lower than, that of the surface code. This, combined with the fact that localised statistics decoding is a parallel algorithm, makes VibeLSD suitable for implementation on specialised hardware for real-time decoding.
Our results establish the colour code as a practical architecture for near-term quantum hardware, providing improved compilation efficiency for both Clifford and non-Clifford gates without incurring additional qubit overhead relative to the surface code.

\end{abstract}

\maketitle

\section{\label{sec:introduction}Introduction}

Quantum error correction (QEC) is key to the manufacture of quantum computing at scale \cite{shor_qec,steane_qec}. The past year has marked a decisive turning point in experimental progress: the first surface code logical qubit operating below the break-even threshold was demonstrated in superconducting hardware \cite{acharya_quantum_2024}, and several experiments have implemented logical gate primitives including lattice surgery \cite{walraff_lattice_surgery,lacroix_scaling_2025}, magic-state injection \cite{Gupta_2024}, and code switching \cite{pogorelov2025experimental,quantinuum_switching,bluvstein_architectural_2025}.

\begin{figure}[!b]
    \centering
    \input{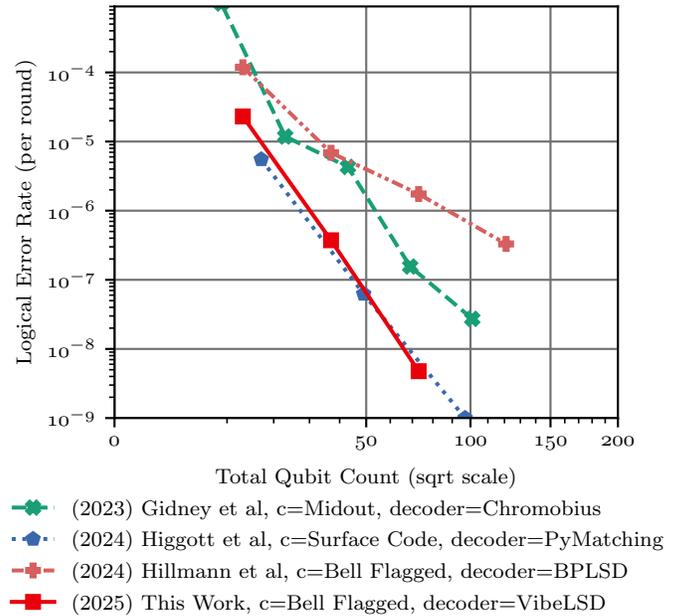}    
    \caption[short]{\justifying 
    Qubit footprint required to reach a target logical error rate at a physical error rate of $10^{-4}$ under the Si1000 noise model. The VibeLSD decoder demonstrates that colour codes with \textit{Bell-flagged} syndrome extraction \cite{baireuther_neural_2019} achieve an overhead nearly identical to that of the surface code decoded with PyMatching \cite{higgott_pymatching_2022}. This represents a substantial improvement over the previous state-of-the-art Chromobius decoder for colour codes \cite{gidney_new_2023}.}
    \label{fig:comparison}
\end{figure}

Selecting a QEC protocol is a critical architectural decision, requiring trade-off between code parameters---such as encoding rate, distance, and logical compilation efficiency---and hardware constraints including noise characteristics, connectivity, and gate fidelities. Equally important is the availability of an efficient decoder, a classical co-processor tasked with processing syndrome measurements in real time. The capabilities of the decoder ultimately constrain QEC performance: its speed determines logical clock rates, while its accuracy governs the qubit overhead needed to achieve reliable computation \cite{terhal_backlog}.

The surface code has long been the industry standard for experimental implementations \cite{fowler_surface_2012, litinski_game_2019}. It benefits from a simple embedding in a degree-four two-dimensional lattice, high thresholds, and efficient decoders based on well-established graph-matching algorithms \cite{higgott_sparse_2025, yue_fusion_blossom, higgott_pymatching_2022,  fowler_minimum_2014}. These features make it the natural choice for the first demonstrations of logical qubits and quantum memories. However, it is less clear whether its advantages will persist as the field advances from quantum memories to computation. A key limitation is the inefficient realisation of common single-qubit logical gates. For example, both the Hadamard and S-phase gates require multiple rounds of stabiliser measurements \cite{Geher2024errorcorrected}. This overhead reduces compilation efficiency and slows down computation at the logical level, raising questions about the long-term suitability of the surface code for fault-tolerant quantum computing.

The colour code is a natural alternative to the surface code, with a local embedding in a two-dimensional lattice \cite{bombin2013introductiontopologicalquantumcodes}. Its principal advantage lies in logical compilation efficiency: the single qubit Clifford gates
$H$ and $S$  can be implemented transversally, with no additional qubit or time overhead, while CNOT gates can be implemented with efficient lattice surgery methods \cite{thomsen_low-overhead_2024, landahl_quantum_2014, fowler_two-dimensional_2011, landahl_fault-tolerant_2011, herzog2025latticesurgerycompilationsurface}. In addition, colour codes form the basis of the recently proposed magic state cultivation protocol, which reduces the cost of non-Clifford gate implementation by an order of magnitude compared with conventional magic state distillation \cite{gidney_magic_2024, lee_low-overhead_2025}.

Despite their appealing features, colour codes have not been viewed as a practical alternative to the surface code, primarily due to the absence of an efficient decoder. Unlike the surface code, they cannot be decoded using graph-matching algorithms, and existing colour code decoders fail to suppress errors sufficiently. This is not a fundamental limitation: simulations with exponential-time integer-programming decoders suggest that colour codes could outperform the surface code under optimal decoding \cite{lacroix_scaling_2025,beni_tesseract_2025}. Unlocking this potential, however, will require the development of practical polynomial-time decoders that can be implemented on specialised hardware and integrated within quantum computing systems to decode syndromes in real-time.

In this work we introduce a \textit{vibe decoding} method as the first practical decoder to bring colour code performance in line with the surface code. At the core of our protocol is an ensemble of belief-propagation (BP) decoders, a class of algorithms widely used in classical communication standards such as 5G and WiFi \cite{ldpc_5g, ldpc_5g_2}. Each BP instance in the ensemble employs a distinct serial message-passing schedule, breaking the symmetries inherent in the colour code decoding graph and thereby increasing the likelihood that at least one decoder produces a valid solution. To further suppress error rates, when the BP ensemble fails, its soft-information outputs are averaged and passed to a localised statistics decoder (LSD) post-processor \cite{hillmann_localized_2024}. This stage applies a parallel sparse-matrix inversion strategy to break code degeneracies and guarantees a solution consistent with the measured syndrome. For brevity, we refer to the full protocol---combining the serial-schedule BP ensemble with the LSD post-processor---as the \textit{VibeLSD} decoder.

We benchmark the VibeLSD decoder against leading colour code decoders using a range of syndrome extraction circuits, error models, and physical error rates. Our central result is summarised in Fig.~\ref{fig:comparison}, which shows the qubit footprint---the number of physical qubits required to reach a target logical error rate---for QEC codes simulated under the SI1000 noise model which is representative of superconducting qubit hardware \cite{eickbusch_demonstrating_2025, gidney_fault-tolerant_2021, gidney_yoked_2025}. The figure demonstrates that colour codes decoded with VibeLSD achieve a qubit footprint nearly identical to that of the surface code decoded with a matching-based decoder, representing a substantial improvement over the previous state-of-the-art Chromobius decoder \cite{gidney_new_2023}. Beyond this, our simulations show that VibeLSD consistently outperforms all existing polynomial-time colour code decoders across a variety of syndrome extraction circuits, noise models, and error rates.

A key strength of VibeLSD is its versatility: it can be applied to any sparse parity-check matrix. This makes it suitable both for decoding quantum memories and for logical gate implementations, where the decoding graph dynamically evolves to capture correlated errors \cite{Sahay_transversal, cain2025fastcorrelateddecodingtransversal, serraperalta2025decodingtransversalcliffordgates}. The decoder also benefits from a straightforward configuration: the ensemble is prepared offline by generating distinct schedules for each constituent decoder. This simplicity stands in contrast to alternative approaches that often require training or extensive hyperparameter optimisation \cite{baireuther_neural_2019, bausch_learning_2024, Varbanov_2025}, offering a practical advantage for large-scale deployment.

The first stage of VibeLSD employs the standard minimum-sum variant of BP, which can be implemented cost-effectively on commercially available FPGA or ASIC hardware \cite{fpga_serial, fpga_flexible, fpga_ldpc_thesis, fpga_hardware,garciaherrero2025diversitymethodsimprovingconvergence}. The LSD post-processor is explicitly designed for parallelisation \cite{hillmann_localized_2024} and is equally well suited to specialised hardware implementations.  The worst-case runtime of VibeLSD is $O(n^3)$, where $n$ is the number of columns in the parity check matrix, placing it on equal footing with matching-based decoders. In practice, however, as with matching decoders  \cite{yue_fusion_blossom, higgott_sparse_2025}, the average runtime is significantly lower, typically sub-linear in the noise regime below threshold.

This paper is structured as follows.
In Section \ref{sec:background} we cover background material on colour codes and their existing decoders.
In Section \ref{sec:SEBPLSD_description} we describe the VibeLSD algorithm and its components.
In Section \ref{sec:benchmarking} we benchmark the new decoder against various circuit-level noise colour code decoders and syndrome extraction circuits from the literature.
We conclude in Section \ref{sec:conclusion} with a discussion about the implications of the new decoder and directions for future work.

\section{Background}\label{sec:background}

\subsection{Colour Codes}\label{sec:colour_code_intro}

\begin{figure}[htpb]
    \centering
    \includegraphics[width=0.6\linewidth]{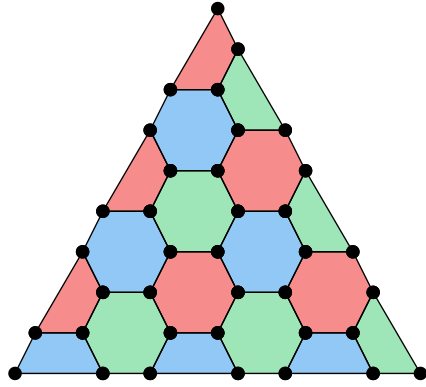}
    \caption[short]{\justifying Distance $7$ hexagonal (6.6.6) tiling colour code.}
    \label{fig:hex_tile}
\end{figure}

Colour codes are Calderbank-Steane-Shor (CSS) quantum stabiliser codes  which encode one logical qubit \cite{bombin_topological_2006}.
The colour codes in this work are defined on a 2D manifold with boundary.
The manifold is cellulated using either a hexagonal (6.6.6) tiling or a square-octagon (4.8.8) tiling. 
The hexagonal tiling uses hexagons in the bulk and quadrilaterals at the boundary (see Fig.~\ref{fig:hex_tile}). 
The 4.8.8 tiling uses hexagons and squares in the bulk and quadrilaterals at the boundary (see Fig.~\ref{fig:488_uni}).  
Both of these tilings are three-colourable---that is, each polygon (plaquette) can be assigned one of three colours (red, green or blue) such that polygons which share an edge are of different colours.

Qubits are associated with the vertices of the lattice and the number of qubits is denoted $n$.
For both tilings, there are  $(n-1)/2$ plaquettes and each of these is associated with two stabiliser generators. 
For a given plaquette, we define an X-type stabiliser generator by applying a Pauli-X operator on each vertex of the plaquette, as well as a Z-type stabiliser generator by applying a Pauli-Z operator on each vertex. 
The stabiliser generators commute because each plaquette has an even number of vertices and the lattice is three-colourable.
The codespace is the simultaneous +1 eigenspace of the stabiliser generators and is of dimension $k=1$ because there are $n$ qubits and $n-1$ independent stabiliser generators.

\begin{figure}[htpb]
    \centering
    \includegraphics[width=0.75\linewidth]{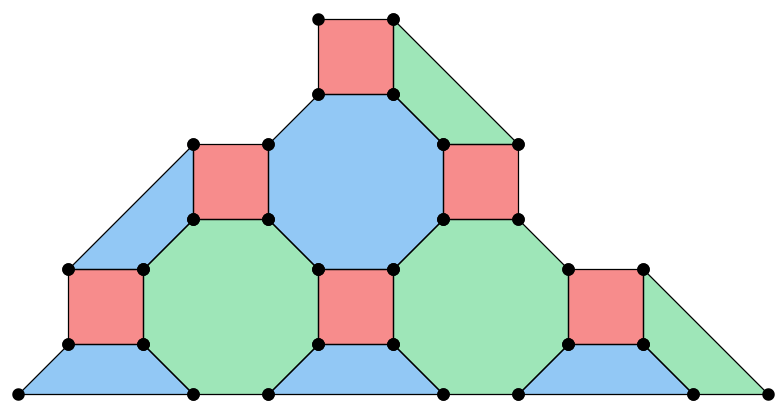}
    \caption[short]{\justifying Distance $7$ 4.8.8 (square-octagon) tiling colour code.}
    \label{fig:oct_tile}
\end{figure}

\subsection{The Decoding Problem}\label{sec:decoding_problem}

Using the symplectic representation, a CSS code can be described by a pair of binary matrices, $H_X$ and $H_Z$, whose rows correspond to the $X$- and $Z$-type stabilisers of the code respectively \cite{gottesman_stabilizer_1997}. Pauli errors are represented as length-$2n$ binary vectors of the form $e = (e_X, e_Z)$. In this framework, the syndrome equation is written as  
\begin{equation}
    s = (s_X, s_Z) = (H_Z e_X \bmod 2, \, H_X e_Z \bmod 2).
\end{equation}
In the \emph{code capacity} noise model, which assumes that only data qubits are affected by errors, decoding reduces to solving $s_X = H_Z e_X \bmod 2$ for $e_X$ and $s_Z = H_X e_Z \bmod 2$ for $e_Z$. In practice, however, stabilisers are extracted via noisy measurement circuits, which introduce additional fault mechanisms.

To account for circuit-level faults, a \emph{detector error model} (DEM) matrix is defined. A \textit{detector} is a linear combination of stabiliser measurement outcomes that sums to zero in the absence of noise. In the DEM formalism, each column corresponds to a possible error mechanism, while each row represents a detector~\cite{aaronson_improved_2004,derks_designing_2024,higgott2024practical}.

The syndrome $s$ obtained from stabiliser measurements identifies which which detectors have been violated. In a stabiliser code, two errors that differ by a stabiliser (or detector) act identically on the logical subspace; equivalently, they belong to the same coset of the stabiliser group. The task of decoding is therefore to infer, from a measured syndrome, the most likely error coset rather than the exact physical error. Choosing any representative from the correct coset restores the logical state, whereas choosing from the wrong coset induces a logical error \cite{dennis_topological_2002,terhal_backlog}.

\subsection{Existing 2D Colour Code Decoders}\label{sec:existing_decoders}

\begin{figure*}[htpb]
    \centering
    \includegraphics[width=\linewidth]{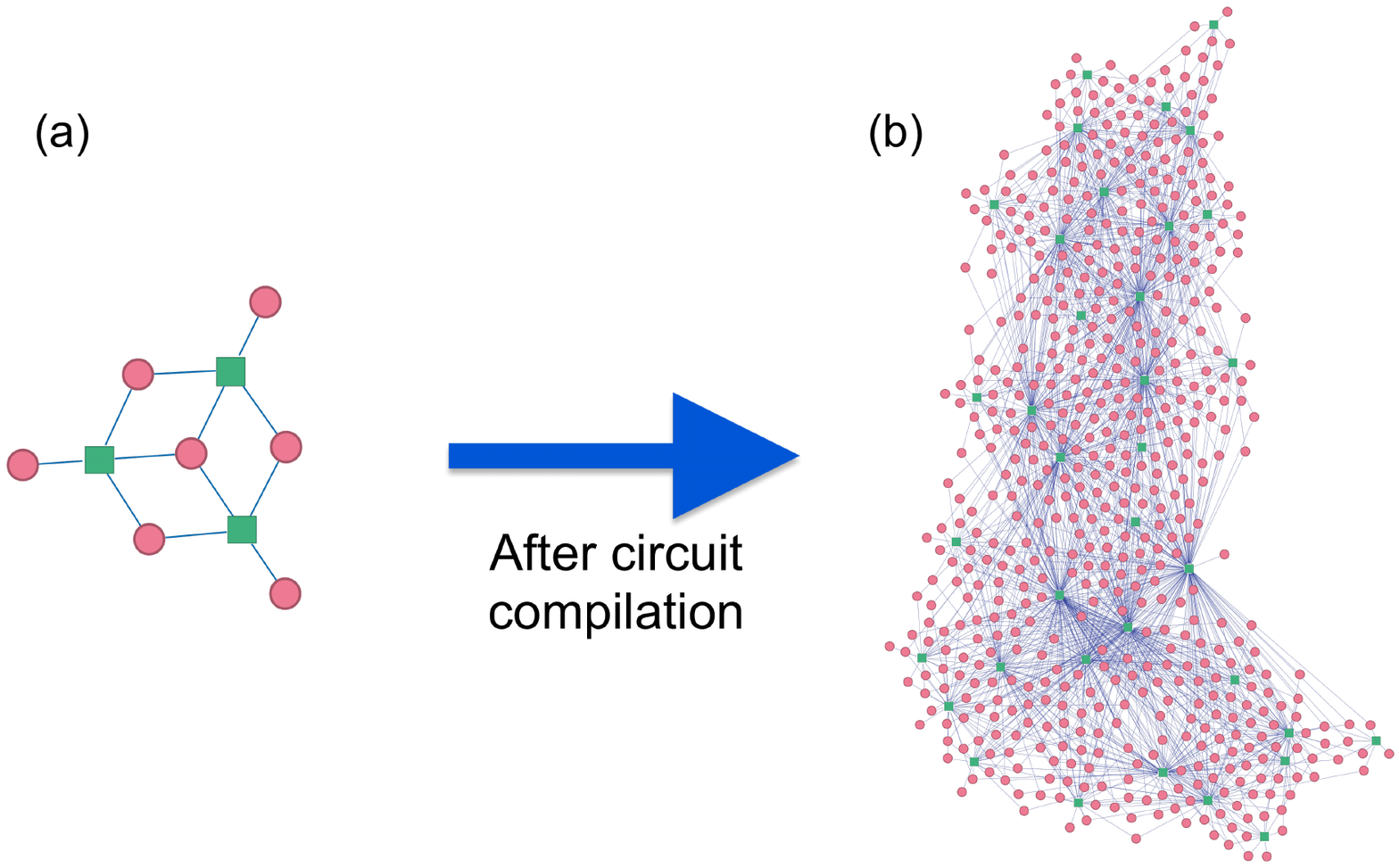}
    \caption[short]{\justifying \textbf{Illustration of Tanner graphs of distance 3 Colour Code: (a) code capacity and (b) Bell-flagged syndrome extraction circuit detector error model}. Green vertices correspond to the detectors and red nodes to the error mechanisms considered for the code capacity error model in (a) and the circuit level noise in (b).}
    \label{fig:tannergraph_colour_d3}
\end{figure*}

To date, most studies of colour code decoding have focused on the code capacity noise model \cite{Duclos_renormal, delfosse_decoding_2014, sarvepalli_efficient_2012, sahay_decoder_2022, berent_decoding_2024, ott_decision-tree_2025}. Circuit-level decoding is more challenging because the DEM matrix is typically much larger and less structured than the original parity check matrix. This is illustrated in Figure~\ref{fig:tannergraph_colour_d3}, which compares the Tanner graph for the $d=3$ colour code code-capacity parity check matrix to the corresponding circuit-level Tanner graph for the DEM matrix.

For surface codes, the DEM matrix is constructed so that circuit faults trigger pairs of syndromes. This allows direct decoding using matching-based algorithms such as union-find \cite{delfosse_almost-linear_2021} or minimum-weight perfect matching \cite{fowler_minimum_2014, higgott_pymatching_2022, higgott_sparse_2025}. In contrast, the colour code DEM matrix is not directly matchable, requiring more general decoding approaches. A common strategy involves decomposing the colour code DEM into subgraphs that are individually matchable, and then combining the results \cite{WangDecoder2010, stephens_efficient_2014, Kubica2023efficientcolorcode, chamberland_triangular_2020, chamberland_topological_2020, gidney_new_2023, lee_color_2025}. Other proposed decoders for colour codes employ neural networks \cite{baireuther_neural_2019} or priority search methods \cite{beni_tesseract_2025}.

In designing decoders for quantum low-density parity-check codes, a common starting point is the belief propagation (BP) decoder \cite{panteleev_degenerate_2021, roffe_decoding_2020}. While standard BP does not typically perform well out-of-the-box for quantum codes, its error rates can be significantly improved using ensemble techniques \cite{koutsioumpas_automorphism_2025, garciaherrero2025diversitymethodsimprovingconvergence} and/or post-processing algorithms \cite{hillmann_localized_2024, iolius2025almostlineartimedecodingalgorithm}. The key advantage of BP-based decoders is their flexibility: they can, in principle, be applied to any sparse parity-check matrix without requiring special properties such as matchability, making them directly applicable to colour code decoding. However, existing BP-based decoders fail to achieve competitive qubit footprints in practice; for example, the BP+LSD decoder shown in Fig.~\ref{fig:comparison} performs worse than the Chromobius decoder.

\section{Vibe Decoding}\label{sec:SEBPLSD_description}

The VibeLSD decoder introduced in this work combines an ensemble of belief propagation decoders operating in parallel with post-processing via localised statistics decoding (LSD). Our approach requires minimal preprocessing, relying only on the generation of a fixed set of random permutations, and is compatible with all colour code syndrome extraction circuits. This flexibility allows us to benchmark different circuits and identify the optimal configuration for each noise model and physical error rate. In this section, we describe the components and configuration of the VibeLSD decoder.

\subsection{Belief Propagation Decoding}\label{sec:BP_decoders}

Belief propagation (BP) is a heuristic decoding algorithm first introduced by Pearl in 1982 \cite{pearl_reverend_1982}. It operates by passing messages between the variable and check nodes of a Tanner graph \cite{tanner_recursive_1981} until either a correction satisfying the syndrome is found or a maximum number of iterations is reached. In classical low-density parity-check (LDPC) codes, BP decoders can achieve performance close to the Shannon limit, the theoretical maximum rate of reliable information transfer over a noisy channel \cite{mackay_near_1996}.

A critical design choice in BP decoding is the \emph{schedule}, which determines the order in which messages are updated across the Tanner graph. The standard approach is a parallel schedule, where all messages are updated simultaneously. However, for codes with highly symmetric graphs---such as colour codes---a \emph{serial} schedule, updating messages sequentially, can dramatically improve performance by breaking symmetries and helping the decoder escape trapping sets \cite{goldberger_serial_2008}.

The VibeLSD decoder leverages this insight by employing the permuted serial BP schedule from \cite{Roffe_LDPC_Python_tools_2022}, based on the \textit{serial-V} schedule of \cite{goldberger_serial_2008}. This choice is particularly effective for colour codes, where graph symmetries otherwise limit the success of standard BP decoders. Pseudo-code for this variant is provided in Algorithm~\ref{alg:bp}.

\subsection{Ensemble Decoding}\label{sec:ensemble_decoding}
The presence of short cycles in the Tanner graph can lead to so-called \textit{split-beliefs} that prevent BP from converging to a solution that satisfies the syndrome equation \cite{richardson2003error, raveendran_trapping_2021}. 
In classical error correction, a method that is used to improve BP convergence is to run multiple instances of the BP algorithm---each differentiated by some aspect of its configuration---in an ensemble \cite{geiselhart_automorphism_2021, krieg_comparative_2025, shen_toward_2025, geiselhart_crc-aided_2020, mandelbaum_endomorphisms,mandelbaum2025subcodeensembledecodinglinear}.

In quantum error correction, short cycles cannot be avoided. On the full stabiliser graph, length four cycles are necessary to ensure stabilisers mutually commute. Furthermore, Tanner graph cycles commonly arise due to degenerate errors \cite{roffe_decoding_2020}. 
To address these issues, ensemble decoders were first explored for surface codes \cite{sheth_neural_2020, shutty_efficient_2024}. 
More recently, papers have extended ensemble methods to the decoding of quantum low-density parity-check (LDPC) codes under circuit-level noise, achieving linear-time complexity solutions \cite{koutsioumpas_automorphism_2025}. 

The VibeLSD decoder we introduce in this work uses an ensemble of BP decoders, each configured with a distinct serial schedule. This approach is inspired by studies in classical error correction that have shown serial schedule ensembling can improve BP performance for codes with short loops \cite{zhang_replica_2005, hussami_polar, geiselhart_crc-aided_2020, krieg_comparative_2025}. In our simulations, we use ensemble sizes of at most $64$, whilst the maximum iteration depth is set to $25$ for all decoders. 

The intuition behind the serial schedule ensemble method is that asynchronous message passing can help break symmetries in the Tanner graph that lead to split beliefs \cite{zhang_shuffled_2005, iterative_poulin_2008, kuo_refined_2020,raveendran_trapping_2021,shen_toward_2025}. This can lead to faster and more reliable convergence of the BP algorithm \cite{goldberger_serial_2008}.

\subsection{Localised Statistics Decoding}\label{sec:lsdecoding}
Belief propagation is often combined with an ordered statistics decoding (OSD) post-processor \cite{Fossorier_osd, roffe_decoding_2020} to handle cases where the BP decoder fails to converge. However, OSD has a worst-case time complexity that scales cubically with the size of the check matrix, limiting its practicality for large-scale experiments.

Localised statistics decoding (LSD) \cite{hillmann_localized_2024} is a recently proposed post-processing algorithm that offers an efficient, parallel alternative to OSD. The key insight behind LSD is that, at low physical error rates, errors tend to be sparse and form multiple small, disjoint decoding problems that can be solved independently. LSD exploits this structure by iteratively growing \textit{clusters} around activated check nodes in the Tanner graph, using the soft information from BP to guide the growth, and merging clusters as necessary to produce a complete correction.

This cluster-based approach yields substantial computational savings. While OSD has a worst-case complexity of $O(m^2 n)$, LSD scales as $O(\text{polylog}(m+n) + \kappa^3)$, where $m$ and $n$ are the numbers of detector and error nodes, respectively, and $\kappa$ is the size of the largest error cluster. The efficiency of LSD is therefore driven by the typically small size of clusters at low physical error rates. We analyse the distribution of cluster sizes in Section~\ref{sec:cluster_analysis}.

\subsection{Description of the VibeLSD Decoder}\label{sec:SEBPLSD_details}

The VibeLSD decoder operates in two stages:
\begin{enumerate}
    \item \textbf{Offline:} preparation of the decoder ensemble (Subsection \ref{offline}).
    \item \textbf{Online:} processing of syndrome data to produce a correction (Subsection \ref{online}).
\end{enumerate}

\subsubsection{Offline Setup of Ensemble Decoders}\label{offline}
The offline setup of the serial-schedule BP ensemble proceeds as follows:
\begin{enumerate}
    \item Select the ensemble size $L$. Larger ensembles generally improve performance and convergence probability, though hardware may limit parallelisation. Empirically, $L=32$ suffices for most circuits up to distance 13.  
    \item Fix a correction limit $M$. Once $M$ BP decoders converge, the remainder are terminated and the best of the $M$ solutions is chosen. We typically use $M=5$, increasing to $M=7$ at higher code distances.
    \item Generate $L$ random permutations $\pi_i$ of $N$ elements, where $N$ is the number of error mechanisms (columns in the circuit-level detector error model parity check matrix).  
    \item Initialise each BP decoder with a schedule permuted by one of the $\pi_i$.  
    \item Initialise an LSD decoder based on the detector error model parity check matrix. We employ the LSD-0 variant, which omits higher-order searches around the initial solution \cite{hillmann_localized_2024}.
\end{enumerate}

Compared to previous methods, the offline stage of VibeLSD is minimal, requiring only the generation of random permutations on $N$ elements.  
The decoder performs reliably across code distances, circuit types, noise models, and error rates without reconfiguration or training.  
In practice, random permutations consistently yield strong BP convergence and overall decoder performance.  
Benchmarking results are presented in Section \ref{sec:benchmarking}.

\subsubsection{Online Ensemble Decoding}\label{online}
\begin{figure}[]
    \centering
    \includegraphics[width=\linewidth]{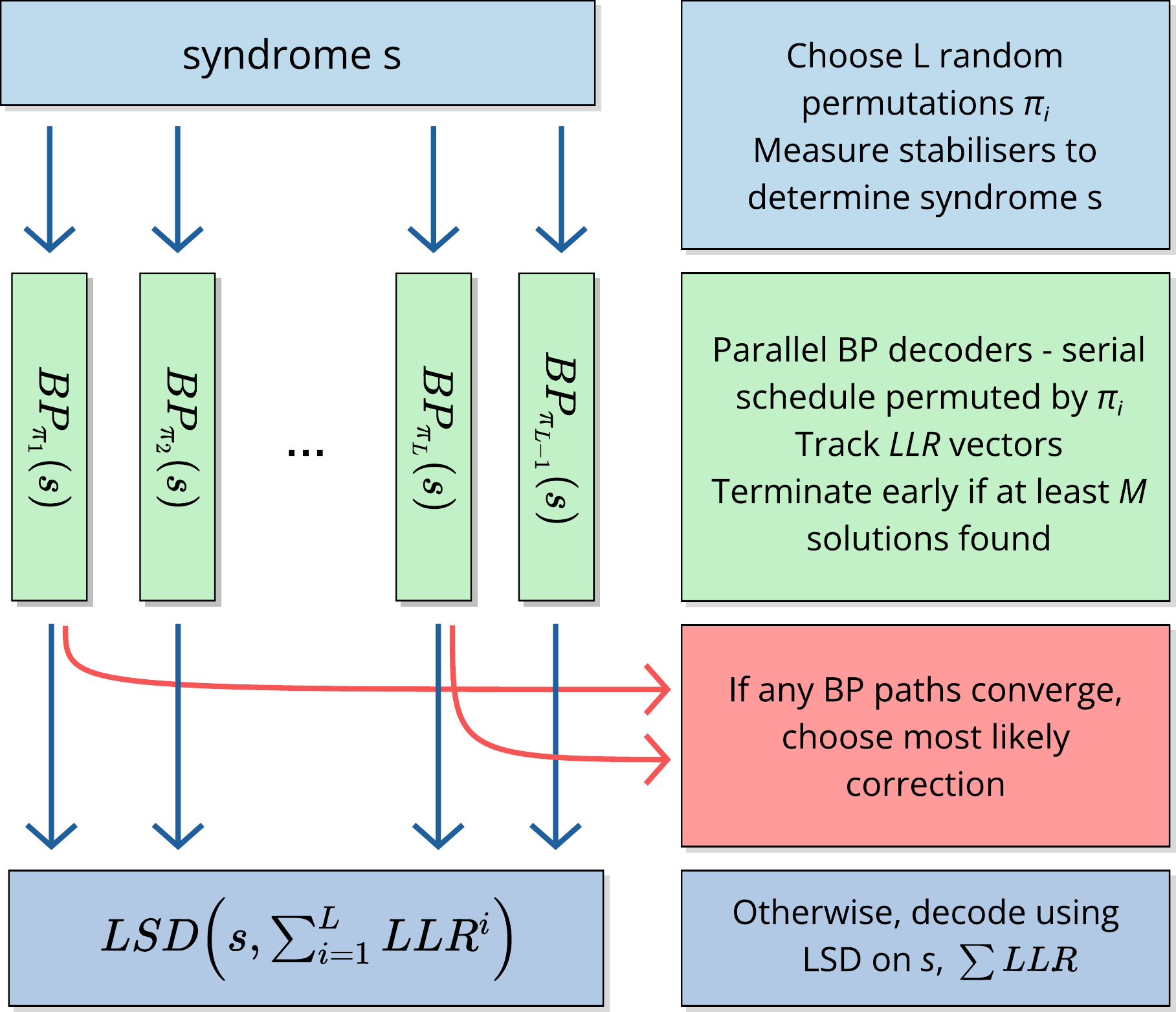}
    \caption[short]{\justifying 
    \textbf{Workflow of the VibeLSD decoder.} 
    In the offline stage, $L$ random permutations are generated to initialise the ensemble of BP decoders. 
    During online decoding, each decoder processes the syndrome using serial-schedule BP updates until convergence or the iteration limit. 
    If at least $M$ decoders converge, the others are terminated and the most likely correction is chosen from the successful candidates. 
    If no decoder converges, the normalised average of their soft outputs is passed to LSD for final decoding.
    }
    \label{fig:SEBPLSD_diagram}
\end{figure}

Once the ensemble is set up, online decoding proceeds as follows:
\begin{enumerate}
    \item\label{llr_list} For each decoder $i$ ($1 \le i \le L$), initialise an empty array $LLR^i$ of length $N$ to store its log-likelihood ratios (LLRs), which are updated during BP iterations. 
    \item Measure stabiliser generators, obtain the syndrome vector $s$, and pass it to all $L$ BP decoders.  
    \item If $M$ decoders converge, terminate the remainder.  
    \item If at least one decoder converges, collect the candidate corrections and rank them using a likelihood metric. In our simulations, this metric is derived from pre-computed priors in the detector error model.  
    \item If no decoder converges, compute the normalised sum of all LLR vectors
    $$
    \frac{1}{L}\sum_{i=1}^L \frac{LLR^i}{\sqrt{\sum_{j=1}^N (LLR_j^i)^2}}
    $$
    and pass it to LSD for decoding. This averages the information across the ensemble.  
\end{enumerate}
A schematic overview of the online decoding process is shown in Figure \ref{fig:SEBPLSD_diagram}.

\subsection{Comparison with Automorphism Ensemble Decoder}

The Automorphism Ensemble (AutDEC) decoder of \cite{koutsioumpas_automorphism_2025} introduces a distinct ensembling strategy for circuit-level noise decoding of Bivariate Bicycle codes \cite{bravyi_high-threshold_2024}.  
Its key idea is to exploit code and circuit automorphisms that map short cycles in the Tanner graph to longer ones, thereby improving BP convergence.  

Numerical results in \cite{koutsioumpas_automorphism_2025} show that AutDEC achieves strong performance on Bivariate Bicycle codes, matching BP+OSD-0 without relying on any post-processing.  
However, when we attempted to apply AutDEC to the colour code, we observed that existing colour code circuits do not have sufficient symmetries to create an ensemble. This motivated the search for alternative ensembling strategies. 

The VibeLSD decoder we introduce here employs a different ensemble strategy. Rather than depending on Tanner graph structure or explicit automorphism groups, VibeLSD simply applies random permutations of error mechanisms across ensemble paths.  
This approach avoids the computational overhead of calculating automorphism groups, which can be challenging for large detector error model (DEM) check matrices \cite{sayginel_fault-tolerant_2024}.  
As a result, we expect VibeLSD to perform robustly across a wide range of quantum error correction codes, including colour codes where AutDEC provides little benefit.

\section{Benchmarking}\label{sec:benchmarking}

We benchmark the VibeLSD decoder against the best known colour code decoders for circuit-level noise \cite{gidney_new_2023, lee_color_2025}.  
Subsection \ref{sec:noise_models} outlines the noise models and error rates considered.  
We then present $X$-memory simulation results for a range of syndrome extraction circuits and circuit-level colour code decoders from the literature, covering codes of distance 3--13. Specifically, we evaluate:  
\begin{itemize}
    \item Bell-flagged circuits with the neural network decoder of \cite{baireuther_neural_2019} under uniform circuit-level noise (Subsection \ref{sec:bell-flagged-circuits});
    \item Tri-optimal circuits with the concatenated MWPM decoder of \cite{lee_color_2025} under uniform circuit-level noise (Subsection \ref{sec:tri-optimal-circuits});
    \item Superdense circuits with the Chromobius decoder \cite{gidney_new_2023,gidney_2023_10375289,noauthor_quantumlibchromobius_2025} under Si1000 circuit-level noise (Subsection \ref{sec:superdense_circuits});
    \item Hexagonal (6.6.6 tiling) and square-octagon (4.8.8 tiling) colour codes with middle-out circuits and the Chromobius decoder under uniform circuit-level noise (Subsection \ref{sec:middle-out-circuits}).
\end{itemize}

For benchmarks up to distance $d=11$, we use an iteration limit of 20 per min-sum BP decoder with an ensemble size of 32.  
For distance-13 superdense and tri-optimal circuits, we increase to 25 iterations and an ensemble size of 64.  
We also use an ensemble of size 64 and an iteration limit of 25 for Bell-flagged and superdense circuits in the footprint comparison of Figure \ref{fig:footprint_all} at $p=10^{-3}$ and $p=10^{-4}$.

Since VibeLSD relies on LSD postprocessing, Subsection \ref{sec:cluster_analysis} analyses how often postprocessing is invoked and the typical complexity of the resulting decoding problem, measured by maximum cluster size.  

We also compare the footprint (the number of qubits required to achieve a target logical error rate) against leading colour code decoders and the surface code with MWPM under $X$-memory circuit-level noise.  
In particular, we show that VibeLSD achieves a footprint comparable to the surface code using Pymatching \cite{higgott_pymatching_2022} (Subsection \ref{sec:footprint_comparisons}).  
We further compare the surface code under independent $X$/$Z$ decoding with the colour code cycling through $X$, $Y$, and $Z$ stabiliser measurements from \cite{gidney_new_2023}, and highlight that VibeLSD outperforms the previous best colour code footprint (Chromobius with midout circuits) in Subsection \ref{sec:xyz_footprint}.

\subsection{Noise models}\label{sec:noise_models}

We consider two noise models in our simulations: the \textit{Uniform} and \textit{Si1000} models.  
The \textit{Uniform} model is a simple baseline commonly used for benchmarking, which facilitates direct comparison with previous works (Table~\ref{table:Uniform_noise}).  
The Si1000 model \cite{eickbusch_demonstrating_2025, gidney_fault-tolerant_2021, gidney_yoked_2025} is derived from the physical characteristics of superconducting hardware (see Table~\ref{table:Si_noise}).  
Superconducting architectures feature 2D planar connectivity, making them naturally suited for implementing 2D colour codes.

\begin{table}[htpb]
    \begin{tabular}{l@{\hskip 10pt}lc} 
    \multicolumn{3}{c}{Si1000} \\ \toprule
    Operation & Error channel & Probability \\ \midrule
    CZ gate & two-qubit depolarising & p \\ 
    Single qubit Cliffords & one-qubit depolarising & p/10 \\
    Initialisation ($|0\rangle$) & bit-flip & 2p \\ 
    Measure (Z basis) & bit-flip & 5p \\ 
    Idling - gates & one-qubit depolarising  & p/10 \\
    Idling - measure, reset & one-qubit depolarising & 2p \\
    \bottomrule
    \end{tabular}
    \caption[short]{\justifying Operations, their associated error channels and probabilities for the Si1000 noise model, which resembles real superconducting hardware.}
    \label{table:Si_noise}
\end{table}

\begin{table}[htpb]
    \begin{tabular}{l@{\hskip 10pt}lc}
    \multicolumn{3}{c}{Uniform} \\ \toprule
    Operation & Error channel & Probability \\ \midrule
    Two qubit Cliffords & two-qubit depolarising & p \\ 
    One qubit Clifford & one-qubit depolarising & p \\
    Reset in X (Z) basis & phase (bit) -flip & p \\
    Measure in X (Z) basis & phase (bit) -flip & p \\
    \bottomrule
    \end{tabular}
    \caption[short]{\justifying Operations, their associated error channels and probabilities for the Uniform noise model.}
    \label{table:Uniform_noise}

\end{table}

\subsection{Bell-Flagged Circuits}\label{sec:bell-flagged-circuits}

Bell-flagged syndrome extraction circuits for the hexagonal colour code are introduced in \cite{baireuther_neural_2019}.  
Each stabiliser employs one flag auxiliary qubit to detect potentially damaging ``hook'' errors.  

\begin{figure}[htpb]
       \begin{center}
    \input{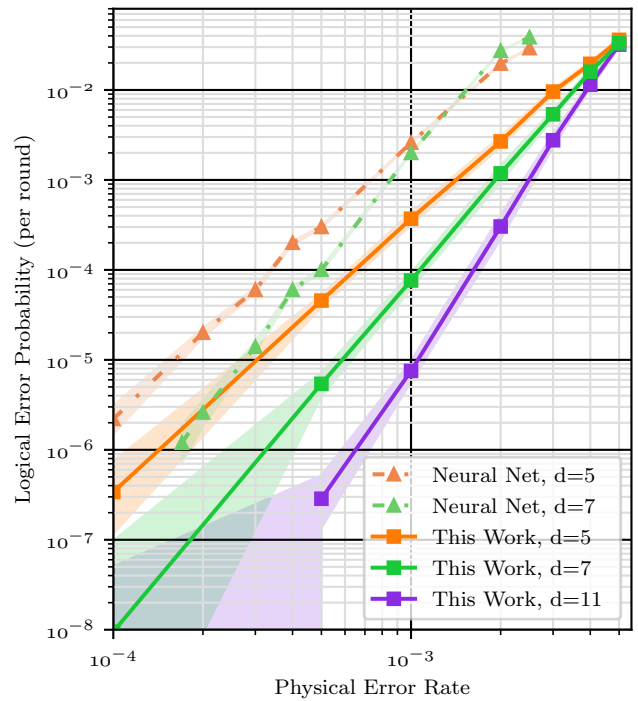}
       \end{center}
       \caption[short]{\justifying \textbf{Bell-flagged circuit results under Uniform circuit-level noise.}  
       Dash-dotted lines show the logical error rate of the neural network decoder from \cite{baireuther_neural_2019}, and solid lines show VibeLSD.  
       Circuits of the same distance share the same colour. Shading indicates hypotheses within a factor of $1,000$ of the maximum likelihood hypothesis based on the sampled data.}
    \label{fig:bellflag_uni}
\end{figure}

The same work also introduces a neural network decoder.  
Figure~\ref{fig:bellflag_uni} compares VibeLSD with this neural network decoder under the uniform noise model.  
Across all physical error rates and code distances, VibeLSD achieves lower logical error rates.

\subsection{Tri-optimal Circuits}\label{sec:tri-optimal-circuits}

\begin{figure}[!t]
    \begin{center}
        \input{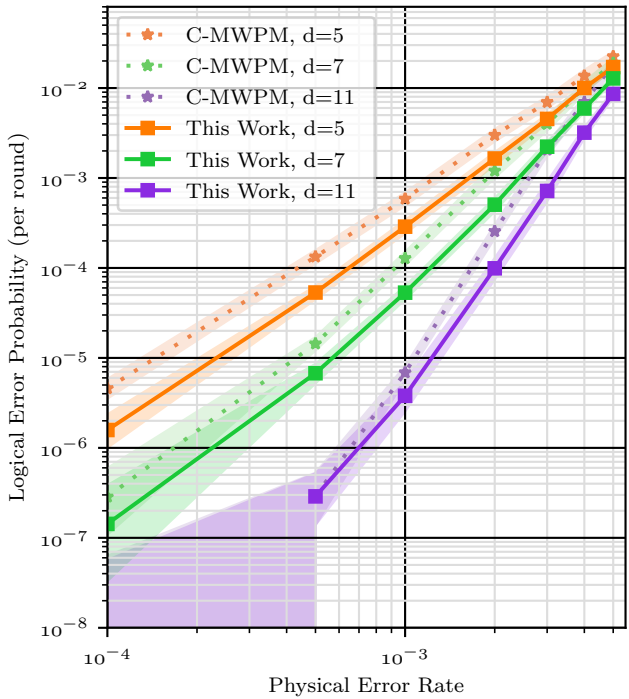}
    \end{center}
    \caption[short]{\justifying \textbf{Tri-optimal circuit results under uniform circuit-level noise.}  
    Dotted lines show the logical error rate of the concatenated MWPM decoder \cite{lee_color_2025}, while solid lines correspond to VibeLSD (this work).}
    \label{fig:triopt_res}
\end{figure}

In \cite{lee_color_2025}, circuits using one physical auxiliary qubit per stabiliser are categorised into equivalence classes.  
The authors employ a \textit{concatenated} minimum-weight perfect-matching (MWPM) approach to identify the circuit with the lowest logical error rate, termed \textit{Tri-optimal}.  
Figure~\ref{fig:triopt_res} compares the logical error rate of VibeLSD with the concatenated MWPM decoder under uniform circuit-level noise.  

\subsection{Superdense Circuits}\label{sec:superdense_circuits}

\begin{figure}[!t]
    \begin{center}
        \input{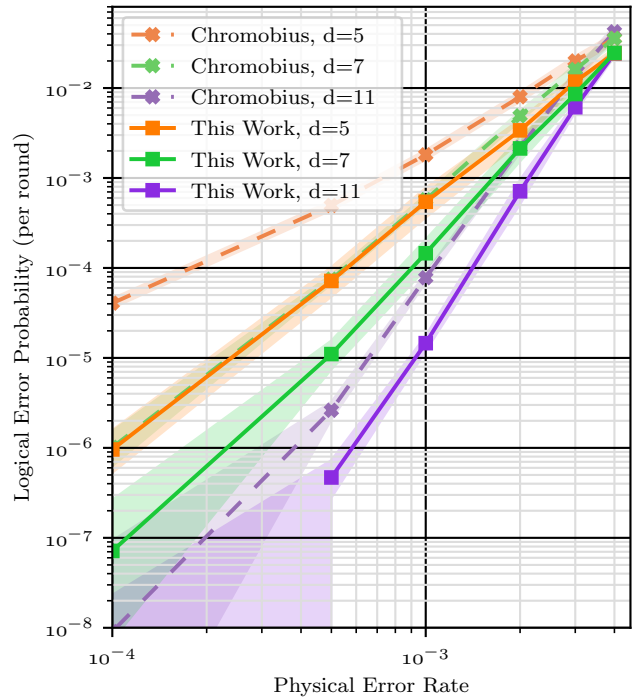}
    \end{center}
    \caption[short]{\justifying \textbf{Superdense circuit results for Si1000 circuit-level noise.}  
    Dashed lines show the logical error rate of Chromobius \cite{gidney_new_2023}, while solid lines correspond to VibeLSD (this work).}
    \label{fig:superdense_comparison}
\end{figure}

The superdense circuits of \cite{gidney_new_2023} provide a more compact variation of the Bell-flagged circuits \cite{baireuther_neural_2019}.  
They employ \textit{Bell multiplexing} and concepts from superdense coding \cite{bennett_communication_1992}, using a pair of auxiliary qubits prepared in a Bell state to simultaneously accumulate the outcomes of both $X$-type and $Z$-type stabilisers.  
Figure~\ref{fig:superdense_comparison} compares VibeLSD with Chromobius under the Si1000 noise model.  

\begin{figure}[!t]
    \begin{center}
        \input{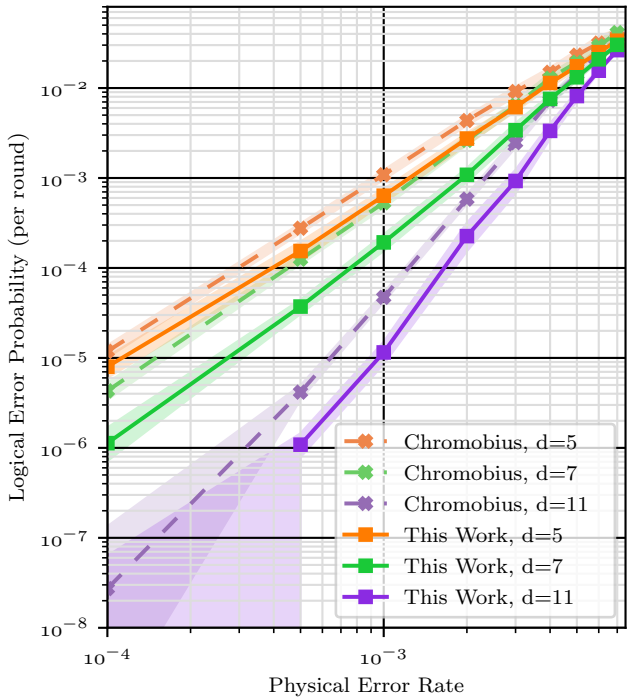}
    \end{center}
    \caption[short]{\justifying \textbf{Hexagonal Midout circuit results under Uniform noise.}  
    Dashed lines show the logical error rate of Chromobius \cite{gidney_new_2023}, while solid lines correspond to VibeLSD (this work).}
    \label{fig:midout_comparison}
\end{figure}

\begin{figure}[!t]    
    \begin{center}
        \input{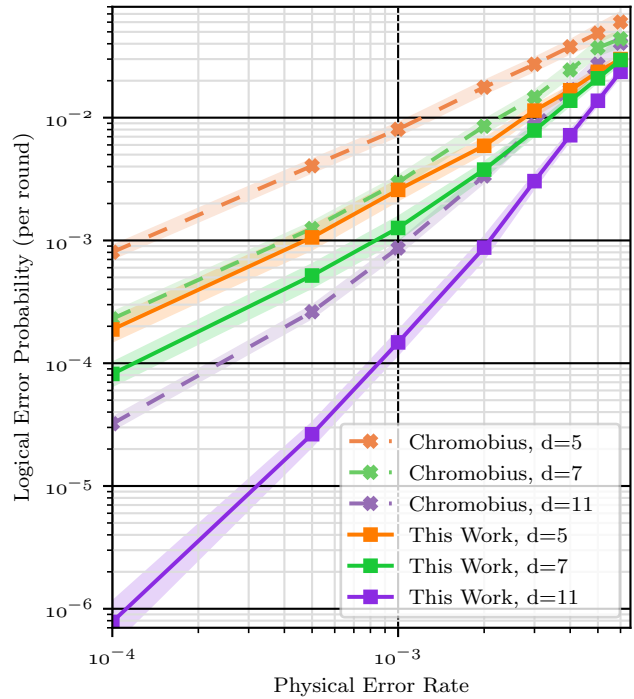}
    \end{center}
    \caption[short]{\justifying \textbf{Square-octagon (4.8.8) Midout circuit results under Uniform circuit-level noise.}  
    Dashed lines show the logical error rate of Chromobius \cite{gidney_new_2023}, while solid lines correspond to VibeLSD (this work).}
    \label{fig:488_uni}
\end{figure}

\subsection{Middle-out Circuits}\label{sec:middle-out-circuits}

The \textit{middle-out} (Midout) circuits for colour codes are introduced in \cite{gidney_new_2023} and are based on earlier constructions for surface codes \cite{mcewen_relaxing_2023}.  
Rather than adding extra auxiliary qubits, the parity of each stabiliser is \textit{folded} onto a data qubit for measurement, reducing both the space overhead and the number of two-qubit gates.  
This approach allows potential hook errors, which halve the circuit distance, but the compactness maintains strong performance in physical error rate regimes of interest, such as $10^{-3}$.  Figures~\ref{fig:midout_comparison} and \ref{fig:488_uni} compare VibeLSD with Chromobius \cite{gidney_2023_10375289, noauthor_quantumlibchromobius_2025} for the hexagonal (6.6.6) and square-octagon (4.8.8) tilings under uniform circuit-level noise.  

\begin{figure*}[htpb]
           \resizebox{\linewidth}{!}{\input{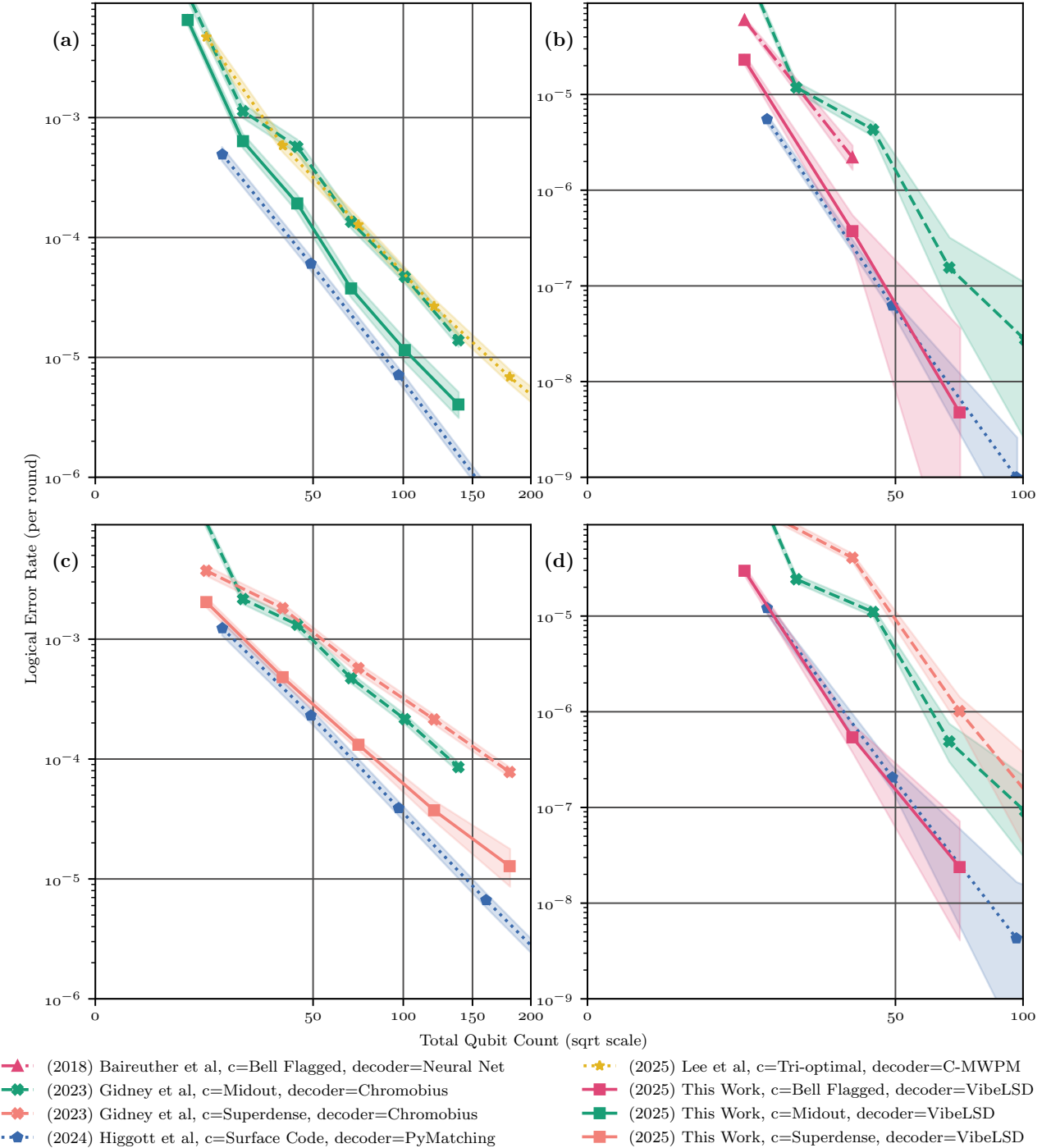}}
        \caption[short]{\justifying \textbf{Footprint comparisons for $X$ memory.}  
Panels (a)–-(d) correspond to: (a) uniform circuit-level noise at $p=10^{-3}$, (b) uniform noise at $p=10^{-4}$, (c) Si1000 noise at $p=10^{-3}$, and (d) Si1000 noise at $p=10^{-4}$.  
In each case, we compare VibeLSD (solid lines, showing the best-performing circuit) against previous state-of-the-art colour code decoders (dashed lines) and the surface code (dotted line). Circuits of the same type share the same hue.  
Results from Baireuther et al.\ are adapted from Figure 4 of \cite{baireuther_neural_2019}, adjusted to per-round rather than per-circuit-layer using the analysis of \cite{gidney_new_2023, noauthor_quantumlibchromobius_2025}. While the comparison is reasonably accurate, slight differences in uniform noise models and the estimation of data points from published plots may introduce minor discrepancies.  
Highlighted regions indicate hypotheses within a factor of 100 of the maximum likelihood hypothesis, given the sampled data.}

        \label{fig:footprint_all}
    \end{figure*}

\subsection{Footprint Comparisons}\label{sec:footprint_comparisons}

We analyse the footprints of each circuit/decoder combination, defined as the number of physical qubits required to achieve a target logical error rate.  
Figure~\ref{fig:footprint_all} compares the footprint of VibeLSD with those of the circuit/decoder combinations in \cite{baireuther_neural_2019, gidney_new_2023, lee_color_2025} under both the uniform depolarising and Si1000 circuit-level noise models at error rates $p=10^{-3}$ and $p=10^{-4}$.  
For reference, we also include the corresponding surface code footprint using PyMatching \cite{higgott_sparse_2025, higgott_pymatching_2022}.  

The results show that VibeLSD achieves a lower footprint than all previous circuit-level colour code decoders and is comparable to the surface code with a matching decoder.  
The optimal circuit depends on the noise model and error rate.  
For the uniform noise model, different circuits perform best at different error rates.  
Under the Si1000 model, superdense circuits outperform midout circuits at high physical error rates, while Bell-flagged circuits perform better at low error rates.  
Across all scenarios, VibeLSD maintains a footprint close to that of the surface code, albeit with the best-performing circuit varying by regime.  Additional plots, including estimated data from \cite{chamberland_triangular_2020}, appear in Appendix~\ref{app:footprint_plots}.

\subsection{XYZ Colour Code Footprint}\label{sec:xyz_footprint}

The $XYZ$ colour code circuits are introduced in \cite{gidney_new_2023}.  
This variant exploits the self-dual structure of colour codes, where multiplying an $X$ and $Z$ stabiliser with identical support produces a corresponding $Y$ stabiliser.  
The $XYZ$ circuit cycles through measurements of the $X$, $Y$, and $Z$ stabilisers on each auxiliary qubit during every syndrome extraction round, providing additional cross-checks that improve sensitivity to correlations introduced by depolarising noise across the three bases.  

Previous polynomial-time decoders struggle in this setting.  
Matching-based decoders, such as Chromobius \cite{gidney_new_2023}, cannot reliably decompose detectors in the $Y$ subgraph, while general-purpose approaches like BP+OSD or BP+LSD perform poorly due to unavoidable short cycles arising from overlapping $X$, $Y$, and $Z$ detectors.  
Consequently, no prior colour code decoder achieves competitive performance for these circuits.  

\begin{figure}[htpb]
    
       \begin{center}
\input{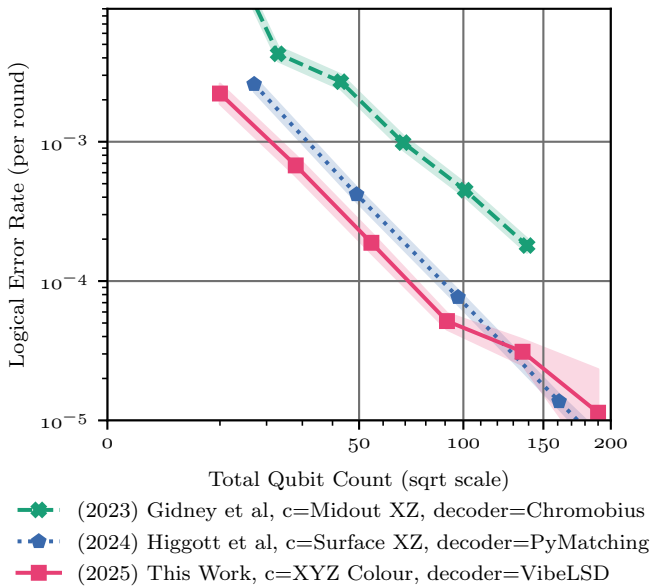}
       \end{center}

        \caption[short]{\justifying \textbf{XYZ colour code footprint for $p=10^{-3}$ Si1000 noise.}  
Green dashed lines show the logical error rate of the Chromobius decoder with independent $X$ and $Z$ decoding of the midout circuit \cite{gidney_new_2023}, and blue dotted lines indicate the surface code decoded independently in $X$ and $Z$ using PyMatching \cite{higgott_pymatching_2022, higgott_sparse_2025}.  
The solid pink line shows the logical error rate of VibeLSD (this work) on the full detector error model of the XYZ colour code.  
VibeLSD outperforms both the Chromobius colour code and the surface code, demonstrating, for the first time, a colour code achieving lower logical error rates than the surface code under practical circuit-level decoding.}
        \label{fig:xyz_footprint}
    \end{figure}

In contrast, VibeLSD performs robustly on the $XYZ$ colour code and, for the first time, surpasses the surface code decoded with PyMatching.  
Figure~\ref{fig:xyz_footprint} shows the decoding footprint of the $XYZ$ circuit under the full detector error model with VibeLSD for the Si1000 noise model at $p=10^{-3}$.  
For reference, we compare against the surface code decoded independently in the $X$ and $Z$ bases; while correlated decoding of the surface code can provide further improvement, it incurs additional overhead \cite{fowler2013optimalcomplexitycorrectioncorrelated}.  

This result demonstrates the versatility of VibeLSD: it succeeds where previous colour code decoders fail and establishes the first example of a colour code outperforming the surface code under practical, circuit-level decoding.

\subsection{LSD Cluster size analysis}\label{sec:cluster_analysis}

In this section, we analyse the frequency with which VibeLSD invokes LSD postprocessing and the typical complexity of the decoding problems it sends to LSD.  
We compare these metrics against the standard BP+LSD-0 decoder without ensembling.  

\begin{figure}[htpb]
    \centering
    \includegraphics[width=\linewidth]{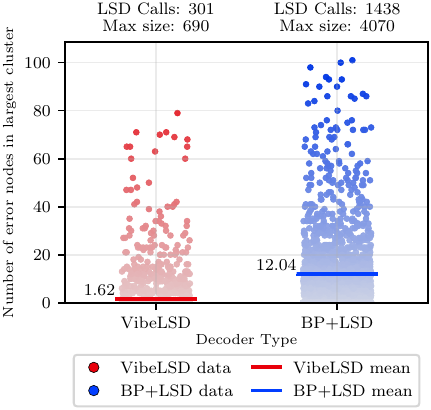}
    \caption[short]{\justifying \textbf{Largest-cluster size distribution and LSD calls for VibeLSD and BP+LSD decoders.}  
Each point represents a single decoding shot, plotted in random order along the $x$-axis within each decoder column.  
The $y$-value indicates the number of error nodes in the largest cluster; points on $y=0$ correspond to shots where the decoder converges without invoking LSD.  
Points are colour-coded by decoder (VibeLSD: red; BP+LSD: blue), with saturation increasing with cluster size (darker = larger clusters).  
Horizontal coloured bars indicate the mean largest-cluster size over $5000$ shots.  
For clarity, nonzero values above the 95th percentile are omitted (VibeLSD: $15$; BP+LSD: $72$).  
The total number of LSD calls and maximum observed cluster size across all shots are reported at the top.}

    \label{fig:cluster_stats}
\end{figure}

Figure~\ref{fig:cluster_stats} presents statistics on the number of LSD calls over $5\,000$ decoding shots for distance-9 colour codes at a physical error rate of $10^{-3}$ using the superdense circuit under Si1000 noise.  
VibeLSD requires significantly fewer LSD calls than BP+LSD-0, with 301 calls over 5000 shots compared to 1438 for BP+LSD-0.  

We also examine the size of the largest cluster formed by LSD, which determines the processing time per call.  
Figure~\ref{fig:cluster_stats} shows that VibeLSD produces cluster sizes that are more tightly distributed and skewed towards smaller sizes compared to BP+LSD-0.  
Additionally, the number of data points above the 95th percentile is substantially lower for VibeLSD.  

These results indicate that VibeLSD not only invokes LSD postprocessing less frequently than BP+LSD-0, but also reduces the computational resources required per call.

\section{Conclusion and Future Work}\label{sec:conclusion}

In this work, we introduce VibeLSD as the first practical decoding algorithm that enables the colour code to match the performance of the surface code under circuit-level noise.  
Our simulations demonstrate that VibeLSD outperforms all existing polynomial-time colour code decoders across a range of syndrome extraction schemes and noise models for error rates below threshold. We identify the optimal circuit for each error rate and noise model that allows matching or outperforming the corresponding surface code footprint.  
Notably, the XYZ variant of the colour code, when decoded with VibeLSD, surpasses the surface code in logical error rate, a first for a practical, polynomial-time decoder under circuit-level noise.  

VibeLSD combines two subroutines---the minimum-sum BP decoder and the LSD postprocessor---both of which can be implemented in parallel.  
This makes the decoder well suited for specialised hardware implementation, such as FPGAs and ASICs, enabling close integration with near-term experimental devices.

The VibeLSD decoder is extremely versatile: our results indicate that it performs well for different colour code circuits at various error rates without needing to be individually tailored or trained. In particular, this makes it a promising decoding approach for logical gate circuits \cite{serraperalta2025decodingtransversalcliffordgates, cain2025fastcorrelateddecodingtransversal} where the decoding graph changes as the computation progresses to account for correlated errors.

In future work, it would be valuable to explore combining different heuristics across the ensemble \cite{iterative_poulin_2008}, as well as alternative post-processing strategies such as those proposed in \cite{lee_color_2025, iolius2025almostlineartimedecodingalgorithm}.  
A limitation of using serial BP schedules is the loss of parallelism at the level of individual decoders within the ensemble. To mitigate this, it would be interesting to investigate layered schedules \cite{crest2023layereddecodingquantumldpc}, which provide a compromise between fully serial and fully parallel update schemes, potentially recovering some parallel efficiency while retaining the convergence advantages of serial schedules.

Although our focus here has been on improving decoding performance to reduce the footprint, the insights gained could also inform the design of better colour code circuits, as highlighted in \cite{gidney_new_2023}. At uniform noise rates near $10^{-3}$, compactness becomes more important than circuit distance, and techniques that reduce the number of two-qubit gates can further decrease the footprint \cite{webster2025heuristicoptimalsynthesiscnot}. In the same spirit, the VibeLSD framework may also extend naturally to other fault-tolerant protocols, including families of high-rate LDPC codes \cite{lin2025singleshottwoshotdecodinggeneralized, jacob2025singleshotdecodingfaulttolerantgates}.

The VibeLSD decoder positions colour codes as a practical alternative to the surface code across all levels of the fault-tolerant stack. Deploying colour codes for fault-tolerant registers could substantially reduce the number of cycles required to execute Clifford gates \cite{thomsen_low-overhead_2024}, leading to significantly lower resource estimates for large-scale quantum computations \cite{gidney_how_2025}. Moreover, colour codes already provide the foundation for low-overhead schemes to prepare non-Clifford states \cite{gidney_magic_2024,gidney_magic_2024}. An important direction for future work is to investigate whether the improved decoding performance of VibeLSD can further reduce the resource costs of fault-tolerant algorithms at scale.

The colour code has long been regarded as a theoretically appealing counterpart to the surface code, but impractical in hardware due to the lack of an efficient decoder. Vibe decoding demonstrates that these advantages are not merely theoretical but can be realised in practice. For any two-dimensional local architecture with the required connectivity, the colour code should now be considered a serious option.

\begin{acknowledgments}
The authors thank Adithya Sireesh, Boren Gu, Liam Veeder-Sweeney, Nicholas Fazio, Seok-Hyung Lee, and Dan Browne for insightful discussions.
This work has made use of the resources provided by the Edinburgh Compute and Data Facility \href{http://www.ecdf.ed.ac.uk/}{(ECDF)}.
SK is supported by the Innovate UK project \enquote{QEC Readout Testbed} [reference number 10151107]. Whilst at UCL, SK was supported by the Engineering and Physical Sciences Research Council [grant number EP/Y004620/1 and EP/T001062/1]. TN is supported by the Engineering and Physical Sciences Research Council (grant number EP/W524384/1), the University of Edinburgh and Quantinuum.
HS is supported by the Engineering and Physical Sciences Research Council [grant number EP/S021582/1]. HS also acknowledges support from the National Physical Laboratory. MW is supported by the Engineering and Physical Sciences Research CouncilEP/W032635/1] and [grant number EP/S005021/1]. JR is funded by an EPSRC Quantum Career Acceleration Fellowship (grant code: UKRI1224). JR further acknowledges support from EPSRC grants EP/T001062/1 and EP/X026167/1.
The drawings of colour code lattices and circuits used for simulations are made using the Chromobius \cite{noauthor_quantumlibchromobius_2025} and color-code-stim \cite{lee_color-code-stim_2025} libraries. The drawing of the detector error model graph is made using Gephi \cite{bastian_gephi_2009}.

\end{acknowledgments}

\section{Data availability}

The data this paper's decoding simulations of VibeLSD applied to the colour code can be downloaded from Zenodo: 
\href{https://doi.org/10.5281/zenodo.16919929}{https://doi.org/10.5281/zenodo.16919929} \cite{koutsioumpas_2025_16919930}. 
This dataset contains all numerical results generated in our study, provided in CSV format. 

\bibliography{main}%

\clearpage
\onecolumngrid
\appendix

\section{Belief Propagation Algorithm Description}

This appendix outlines the minimum-sum Belief Propagation (BP) decoding algorithm in Algorithm~\ref{alg:bp}, showing both parallel and serial implementations.  
The serial variant uses the \emph{V-serial} schedule \cite{goldberger_serial_2008}, which sequentially updates messages according to a fixed permutation to break symmetries and improve convergence for highly structured codes such as colour codes.  
In our simulations of the VibeLSD decoder, we use the serial BP implementation from the LDPC Python package of \cite{Roffe_LDPC_Python_tools_2022}.  

Decoding proceeds by iteratively passing messages along the Tanner graph until either the estimated error pattern satisfies the measured syndrome or the maximum number of iterations is reached. The pseudocode illustrates both schedule types, highlighting how the serial schedule differs from the parallel schedule in updating messages immediately as they are computed.

\begin{algorithm}[htpb]
\caption[short]{\justifying Min-Sum Belief Propagation for Syndrome-Based Decoding}
\KwIn{Parity check matrix $H$, syndrome $s$, max iterations $T_{\max}$, Schedule = Parallel or Serial, Permutation $\pi_N \in S_N$ }
\KwOut{Estimated error $\hat{e}$}

\BlankLine
\textbf{Initialization:} For each edge $(v,c)$ between variable node $v$ and check node $c$ in the Tanner graph of $H$,
initialize $m_{v \to c} \leftarrow$ channel log-likelihood ratio (LLR).

\BlankLine
\For{$t = 1$ \KwTo $T_{\max}$}{
    \textbf{Check-to-variable update:} For each check node $c$ and connected $v$:
    \[
    m_{c \to v} \leftarrow \left( \prod_{v' \in N(c) \setminus v} \operatorname{sign}(m_{v' \to c}) \right)
    \cdot \min_{v' \in N(c) \setminus v} |m_{v' \to c}|
    \]
    
    \BlankLine
    \textbf{Variable-to-check update:}
    
    \uIf{\textbf{Schedule = Parallel}}{
        Update all $m_{v \to c}$ simultaneously:
        \[
        m_{v \to c} \leftarrow \lambda_v + \sum_{c' \in N(v) \setminus c} m_{c' \to v}
        \]
    }
    \uElseIf{\textbf{Schedule = Serial}}{
        For each variable node $v$ in order $[\pi(0),\pi(1),...,\pi(N)]$ , update $m_{v \to c}$ immediately
        using the most recent $m_{c' \to v}$ values (including any updated earlier in this iteration).
    }
    
    \BlankLine
    \textbf{Decision:} Compute
    \[
    L_v \leftarrow \lambda_v + \sum_{c \in N(v)} m_{c \to v}
    \]
    and set $\hat{e}_v \leftarrow 1_{[L_v \leq 0]}$.
    
    \If{$H \hat{e} = s$}{
        Return $\hat{e}$\\
        \textbf{break}
    }
}
Return $L_v$
\label{alg:bp}
\end{algorithm}
\newpage
\section{VibeLSD Logical Error Rates for Various Colour Code Circuits and Noise Models}\label{app:SEBPLSD_logical_error_rates}

\begin{figure}[H]
   \input{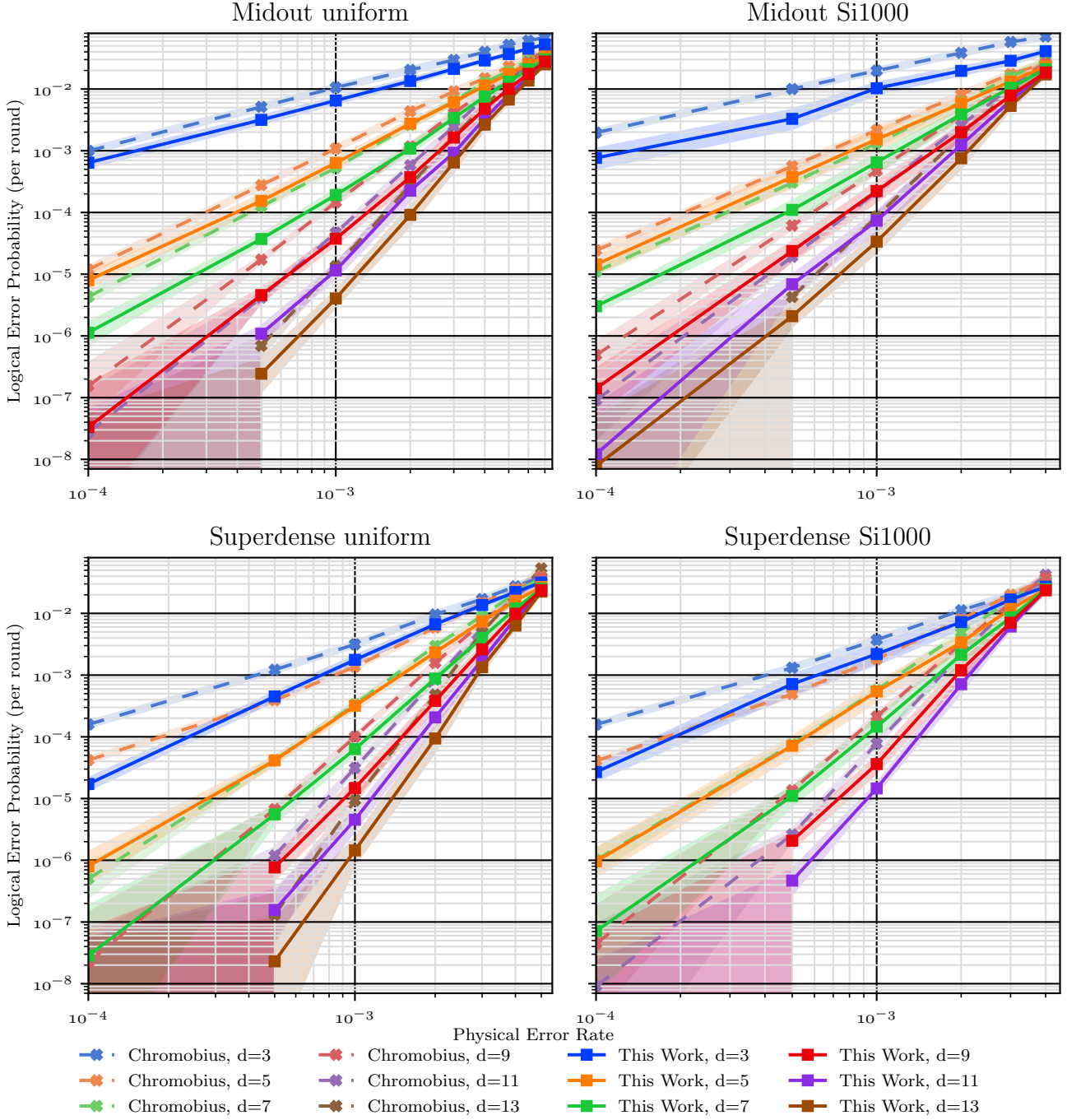}
   \caption[short]{\justifying \textbf{Logical error rate comparison for code distances 3-13 for different circuits and noise models}. Top left: Midout Circuit with Uniform noise. Top right: Midout Circuit with Si1000 noise. Bottom left: Superdense with Uniform noise. 
   Bottom right: Superdense with Si1000 noise. In all plots dashed lines correspond to Chromobius and solid lines to this work.}
   \label{fig:extra_plots}
\end{figure}

In Figure \ref{fig:extra_plots}, we include additional plots of logical error rates for the VibeLSD decoder for various circuit types, noise models, error rates and colour code distances.

\newpage

\section{Additional Footprint Plots}\label{app:footprint_plots}
\begin{figure}[htpb]
    \centering
    \resizebox{\textwidth}{!}{
        \input{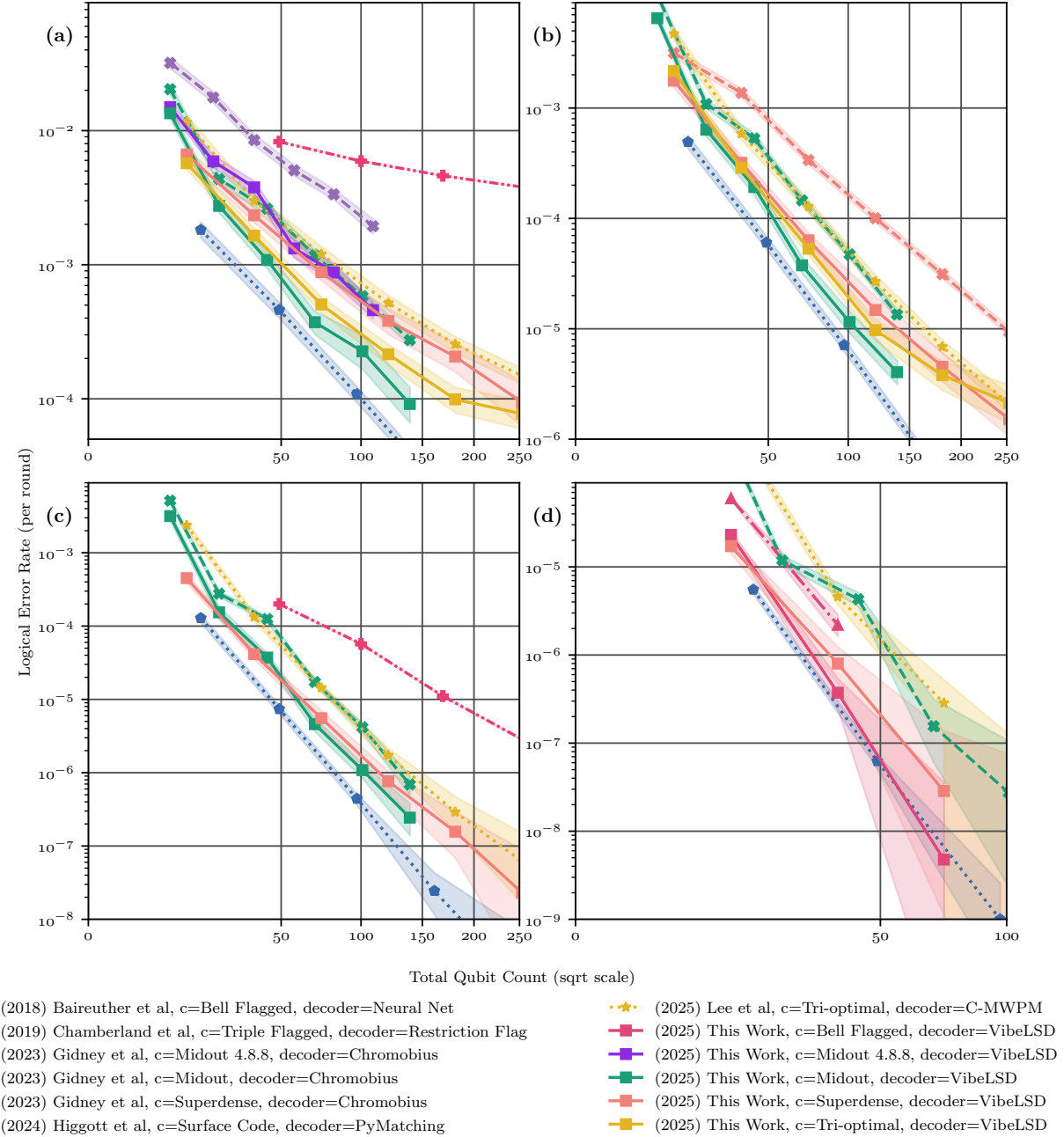}
    }
    \caption[short]{\justifying \textbf{Footprint Comparison - Uniform circuit-level noise}. 
    (a): $p=0.002$. (b): $p=0.001$. (c): $p=0.0005$. (d): $p=0.0001$. In each case we compare against previous decoders (dashed and dotted lines) as well as the surface code (dotted brown line). Circuits of the same type are given the same hue. The results of our decoder are shown in solid lines in each case.}
    \label{fig:more_footprints_uni}
\end{figure}

In Figure \ref{fig:more_footprints_uni} we show footprints for the uniform error model. The results from \enquote{Baireuther et al} are based on Figure 4 of \cite{baireuther_neural_2019} adjusted to be per-round rather than per-circuit-layer. The results from \enquote{Chamberland et al} come from Figure 10(a) of \cite{chamberland_triangular_2020} modified to be per-round instead of per-shot. The comparison is reasonably accurate but not exact due to slight variations in the uniform noise models between the papers and potential inaccuracies because we estimated data points based on their plots due to a lack of raw data.

\end{document}